\documentclass[twocolumn,showpacs,preprintnumbers,amsmath,amssymb]{revtex4}
\begin{document}

\title{Multiple M-wave interaction with fluxes}

\author{
Igor A. Bandos $^{\dagger\ddagger}$}
\address{$^{\dagger}$Department of
Theoretical Physics, University of the Basque Country
 EHU/UPV, P.O. Box 644, 48080 Bilbao, Spain \\ $^{\ddagger}$
IKERBASQUE, Basque Foundation for Science, 48011, Bilbao, Spain}

\date{V1: March 1, 2010. V2: May 14, 2010. {\bf Phys. Rev. Lett. {\bf 105}, 071602 (2010), het-th/1003.0399}}

\def\theequation{\arabic{section}.\arabic{equation}}

\begin{abstract}

We present the equations of motion for multiple M$0$--brane
(multiple M-wave or mM0) system in general eleven dimensional  supergravity
background. These are obtained in the frame of superembedding
approach, but have a rigid structure: they can be restored from
 SO(1,1)$\times\,$SO(9) symmetry characteristic for M$0$.
BPS (Bogomol'nyi--Prasad--Sommerfield) conditions for the 1/2
supersymmetric solution of these equations have the fuzzy 2-sphere
solution describing M2-brane.

\end{abstract}


\maketitle

\renewcommand{\theequation}{\arabic{equation}}

Supersymmetric extended objects, (super-)$p$--branes (including
string for $p$=1, membrane for $p$=2 and also particle for $p$=0) and
interacting systems of several branes play very important r\^ole in
String/M-theory \cite{M-theory} and in the AdS/CFT correspondence
\cite{AdS}. They are used in constructing models of our Universe
as a 3--brane or an intersection of p--branes in the space of
higher--dimensions. Such {\it Brane World} scenarios can be developed in
the frame of string/M-theory as well as independently of it. The
most known examples of the later were the Randall--Sundrum models
\cite{Randall:1999ee} which then were incorporated in the
M-theoretic context in \cite{Duff:2000az}.

The most interesting $p$--branes are $D$=10 fundamental strings and D$p$-branes (Dirichlet $p$-branes), where the fundamental string can have its ends, and $D$=11
 M$p$--branes with $p$=0,2,5. These can be
described by supersymmetric solutions of 10$D$ and 11$D$
supergravity equations (see \cite{Duff94-Stelle98} and refs
therein), by the worldvolume actions
\cite{BST1987,Dpac,B+T=Dpac,blnpst} and in the frame of the
so-called superembedding approach
\cite{bpstv,hs96,hs2,Dima99,IB09:D0,mM0=PLB} (see \cite{Dima99} for more refs).

As far as the multiple p-brane systems are concerned, it was
appreciated long ago that in the  very low energy limit the dynamics
of multiple D$p$--brane (mD$p$) system is approximately described by the maximally
supersymmetric U(N) super--Yang--Mills (SYM) action
\cite{Witten:1995im}. In the search for a counterpart playing the
r\^ole of SYM action for the case of multiple M2-branes, the
Bagger--Lambert-Gustavsson (BLG) model \cite{BLG}, based on the
notion of 3-algebras rather than Lie algebras, and the 3/4
supersymmetric ($d$=3, ${\cal N}$=6)
Aharony--Bergman--Jafferis--Maldacena (ABJM) SU(N)$\times\,$SU(N)
invariant Chern--Simons plus matter model \cite{ABJM} were found.

However in the search for  complete supersymmetric, Lorentz and
diffeomorphism invariant action for multiple D$p$'s (M$p$'s), which
would be a counterpart of the Dirac--Born--Infeld plus Wess--Zumino
action for a single D$p$--brane \cite{Dpac,B+T=Dpac} (single M$p$
\cite{BST1987,B+T=Dpac,bpstv}) only particular progress has been
reached (see \cite{Dima01} for low dimensional and low
co-dimensional branes and \cite{Howe+Linstrom+Linus} and
\cite{IB09:D0} discussed below).

The widely accepted purely bosonic Myers `dielectric brane action'
\cite{Myers:1999ps}, which was generalized for the case of multiple
M-waves (multiple M0--branes or mM$0$-s) in \cite{YLozano+=0207},  does not
possess neither supersymmetry nor 10D Lorentz symmetry, nor complete
diffeomorphism invariance. The boundary fermion
approach of \cite{Howe+Linstrom+Linus} certainly provides a complete,
supersymmetric and Lorentz covariant description of
mD$p$-s, but on `minus one quantization level': the
quantization of the auxiliary boundary fermion variables is needed
to arrive at a description of mD$p$ systems similar
to the description of a single D$p$--brane in \cite{Dpac}.

To search for a (possibly approximate but going beyond the $U(N)$
SYM approximation) Lorentz and diffeomorphism covariant and
supersymmetric equations of the mD$p$-s,
it was proposed \cite{IB09:D0} to use superembedding approach
\cite{bpstv} which had shown its efficiency in searching for the
single D$p$-brane and single M$5$--brane equations \cite{hs96,hs2}. It
was shown in \cite{IB09:D0} that the superembedding approach for the
mD$0$ (multiple D-particle) system results
in selfconsistent dynamical equations for matrix
superfields describing relative motion of the mD0 constituents and
that the structure of the bosonic equations in an arbitrary type IIA
supergravity background shows the Emparan--Myers dielectric brane
effect \cite{Emparan:1997rt}, \cite{Myers:1999ps}. In the case of
flat superspace the mD$0$ equations of  \cite{IB09:D0} coincide with
the result of dimensional reduction of 10D SYM down to $d=1$, which
are the starting equations of the Matrix model of \cite{Banks:1996vh}.
Then,  the question arose whether one could show the restoration
of the 11D Lorentz symmetry in this description of mD$0$, like it
was the case for the Matrix model. The affirmative answer on this
has been given in \cite{mM0=PLB}, where the superembedding
approach for mM$0$ system was developed and used to derive mM$0$ equations of motion in the case of flat target superspace. The equations  describing relative motion of mM$0$ constituents coincide with the ones for the relative motion in mD$0$ system in flat 10D type IIA
superspace \cite{IB09:D0}. This, together with the fact that a single D$0$-brane action can be obtained by
dimensional reduction (dualization) of the action for single M$0$
\cite{B+T=Dpac}, allowed us to conclude that the mM$0$ equations obtained from
superembedding approach give an equivalent form of the mD$0$
equations but with restored 11D Lorentz invariance.

In this letter we present the equations of motion  for mM$0$ system
{\it in curved 11D superspace} which describe the
mM$0$ interaction with the 11D supergravity.
To our best knowledge, this is the first covariant generalization of the
Matrix model equations for the case of nontrivial- and not pure
bosonic- 11D supergravity background.

{\bf 1.}
To fix the basic notion and notation, we begin by a very brief description of {\it superembedding approach to a single M$0$--brane}  in general $11D$ supergravity background. This requires the superfield description of 11D supergravity in terms of supervielbein one forms
${E}^{ {A}}:=
dZ^{ {M}}E_{ {M}}{}^{ {A}}(Z)=
 ({E}^{ {a}}, {E}^{ {\alpha}})$ (with bosonic vectorial form ${E}^{a}$, $a=0,1,\ldots, 9, 10$, and fermionic spinorial form ${E}^{ {\alpha}}$,  ${\alpha}=1,\ldots ,
32$) which satisfy the
set of superspace constraints \cite{CremmerFerrara80BrinkHowe80} of which the most important fixes the from of bosonic torsion 2--form of the curved $11D$ superspace $\Sigma^{(11|32)}$ (see \cite{mM0=PLB} and refs. therein  for details)
\begin{eqnarray}
\label{Ta=11D} & T^{{a}}:= DE^{{a}} =
-i{E}^\alpha \,  {E}^\beta \Gamma^{{a}}_{\alpha\beta}\; . \qquad
\end{eqnarray}
Here $\Gamma^{{a}}_{\alpha\beta}=\Gamma^{{a}}_{\beta\alpha}$ are 11D Dirac matrices and the exterior product of differential forms is assumed (${E}^\alpha {E}^\beta ={E}^\alpha \wedge {E}^\beta$). We have denoted the local coordinates of $\Sigma^{(11|32)}$ by ${Z}^{{ {M}}}=
({x}{}^{ {m}}\, ,
{\theta}^{\check{ {\alpha}}})$ ($\check{\alpha}=1,\ldots ,
32$, $m=0,1,\ldots, 9, 10$).

The standard formulation of M--branes (M$p$-branes  with $p=0,2,5$)
deals with embedding of a purely bosonic worldvolume ${W}^{p+1}$
(worldline ${W}^{1}$ for M$0$-case of \cite{B+T=Dpac}) into the {\it
target superspace} $\Sigma^{(11|32)}$.

The {\it superembedding approach} to M-branes \cite{bpstv,hs2},
following the STV (Sorokin--Tkach--Volkov) approach to
superparticles and superstrings \cite{stv} (see \cite{Dima99} for
review and further refs) describes their dynamics in terms of
embedding of  {\it worldvolume superspace} ${\cal W}^{(p+1|16)}$
with $d=p+1$ bosonic and $16$ fermionic directions into
$\Sigma^{(11|32)}$. This embedding can be described in terms of
coordinate functions $\hat{Z}^{{ {M}}}(\zeta)= (\hat{x}{}^{
{m}}(\zeta)\, , \hat{\theta}^{\check{ {\alpha}}}(\zeta))$, which are
superfields depending on the local coordinates $\zeta^{{\cal M}}$ of
${\cal W }^{(p+1|16)}$,
\begin{eqnarray}
\label{WinS} & {\cal W }^{(p+1|16)}\in \Sigma^{(11|32)} : \quad
Z^{ {M}}= \hat{Z}^{ {M}}(\zeta) \;  . \qquad
\end{eqnarray}
For $p=0$ these are  $\zeta^{{\cal M}}=(\tau ,\eta^{\check{q}})$,
where $\tau$ is proper time and $\eta^{\check{q}}$ are 16 fermionic
coordinates  of the {\it worldline superspace} ${\cal W }^{(1|16)}$,
($\eta^{\check{q}}\eta^{\check{p}}=-
\eta^{\check{p}}\eta^{\check{q}}$, $\check{q}=1,\ldots ,16$).

The {\it superembedding equation} states that the pull--back
$\hat{E}^{{a}}:= d\hat{Z}^{M}(\zeta) E_M^{a}(\hat{Z})$ of the
bosonic supervielbein form ${E}^{{a}}:= d{Z}^{M} E_M^{a}(Z)$ to the
worldvolume superspace has no fermionic projection. For the case of
M$0$--brane it reads
\begin{eqnarray}
\label{SembEq=M0}
  \hat{E}_{+q}{}^{{a}}:= D_{+q}\hat{Z}^M E_M{}^{{a}}(\hat{Z}) = 0\; ,
\qquad
\end{eqnarray}
where  $D_{+q}$ is a fermionic covariant derivative of ${\cal
W}^{(1|16)}$, $q=1,...,16$ is a spinor index of $SO(9)$ and $+$
denotes the `charge' (weight) with respect to the local $SO(1,1)$
group. The only bosonic  covariant derivative of ${\cal W}^{(1|16)}$
is denoted by $D_{\#}$:=$D_{++}$ and supervielbein of ${\cal
W}^{(1|16)}$ by $e^{\cal A}= d\zeta^{{\cal M}} e_{{\cal M}}{}^{\cal A}(\zeta) =
(e^{\#}\; , \; e^{+q})$. Notice that in our notation the upper plus index is equivalent to
the lower minus, and vice-versa, so that one can equivalently write,
for instance, $D_{+q}$=$D^-_q$ and $D_{\#}$=$D^{=}/2$.

The superembedding equation (\ref{SembEq=M0}) is {\it on--shell}  in
the sense that it contains the M$0$-brane equations of motion among
their consequences. We refer to \cite{mM0=PLB} for the explicit form
and the derivation of these equations. For the discussion below we
will need only few details concerning the on--shell geometry of the
worldline superspace ${\cal W}^{(1|16)}$.
In particular, with our conventional constraints (see
\cite{mM0=PLB}) the bosonic torsion two form of ${\cal W}^{(1|16)}$
reads
\begin{eqnarray}
\label{De++=efef} De^{\# }=-2ie^{+ q} \, e^{+q}\; , \qquad
\end{eqnarray}
the Riemann curvature two form of ${\cal W}^{(1|16)}$ vanishes, while
fermionic torsion $De^{+q}$ and curvature of the normal bundle over ${\cal W}^{(1|16)}$ are expressed through the following components
of the pull--backs of the bosonic and fermionic fluxes (field
strengths or curvatures)
\begin{eqnarray} \label{M0:Fluxes}
 & \hat{F}_{\# ijk}:= F^{{a}{b}{c}{d}} (\hat{Z}) u_{{a}}
{}^{=}u_{{b}}{}^{i}u_{{c}}{}^{j}u_{{d}}{}^{k}\; , \qquad  \\
\label{M0:RFlux} & \hat{R}_{\# ij \#}:= R_{dc\;
ba}(\hat{Z})u^{d=}u^{ci} u^{bj}u^{a=} \; ,  \qquad
\\ \label{M0:fFluxe} & \hat{T}_{\#\, i\, +q}
:=T_{{a}{b}}{}^{\alpha}(\hat{Z})\,  v_{{\alpha}q}^{\; -}\,
u_{{a}}^{=}u_{{b}}^i\; .   \qquad
\end{eqnarray}
Here $u_{{a}}^{=}$ and $u_{{b}}^i$ are the auxiliary
moving frame superfields which obey (notice that $u_{{a}}^{\#}\not= $  $u_{{a}\#}$:=$u_{{a}}^{=}/2$)
\begin{eqnarray}\label{u++u++=0}
& u_{ {a}}^{=} u^{ {a}\; =}=0\; , \qquad u_{ {a}}^{\# } u^{ {a}\; \#
}=0\; ,  \qquad    u_{ {a}}^{\; \# } u^{ {a}\; =}= 2\; , \quad
\nonumber \\  \label{uiuj=-}  & u_{ {a}}^{=} u^{ {a}\; i}=0\; ,
\qquad
 u_{{a}}^{\;\#} u^{ {a} i}=0\; , \qquad   u_{ {a}}^{ i} u^{ {a} j}=-\delta^{ij}\; . \quad
\end{eqnarray}
The sixteen 11D spinor superfields $v_{{\beta}q}^{\;
-}=v_{{\beta}q}^{\; -}(\zeta)$  in  (\ref{M0:fFluxe}) are the spinor
moving frame variables which can be considered, roughly speaking, as
square roots of the light--like vector  $u_{ {a}}^{=}$ in the sense
of that (see \cite{mM0=PLB} and refs. therein)
\begin{eqnarray}\label{M0:v-v-=u--}
 v_{q}^- {\Gamma}_{ {a}} v{}_{p}^- = \; u_{ {a}}{}^{=} \delta_{qp}\; , \qquad
 & 2 v_{\alpha}{}_{q}^{-}
 v_{ \beta}{}_{q}^{-}{}= {\Gamma}^{a}_{ {\alpha} {\beta}} u_{ {a}}{}^{=}\; . \qquad
\end{eqnarray}

Notice that the {\it equations of motion for single M$0$--brane} can be
expressed by the statement that $v_{\alpha q}^{\; -}$ and the
light--like moving frame vector $u_a^{=}$ are covariantly
constant,   $Dv_{\alpha q}^{\; -} =0$ and  $Du_a^{=}=0$. Then one finds
\begin{eqnarray} \label{DFijk=}
& D_{+q}\hat{F}_{\# ijk}= 3i \gamma_{[ij|\; qp} \hat{T}_{\#\, |k]\;
p}
   \; ,   \qquad
\\ \label{DpT--iq=} & D_{+p} \hat{T}_{\# \, i\, + q}= {1\over 2}
\hat{R}_{\# ij \#} \gamma^j_{pq} + {1\over 3} D_{\#} \hat{F}_{\#
ijk} \gamma^{jk}_{pq} \qquad \nonumber
\\ & +  {1\over 18} D_{\#} \hat{F}_{\#
jkl} \gamma^{ijkl}_{pq} +   \hat{F}_{\# [3]}\hat{F}_{\# [3']}
{\Sigma}{}^{i\, , \,  [3]\, ,\, [3']}_{pq}
  \; ,    \qquad
\end{eqnarray}
where ${\Sigma}{}^{i,[3],[3']}_{pq}:={\Sigma}{}^{i\, , \,
j_1j_2j_3\, ,\, k_1k_2k_3}_{pq}$ is a constant tensor-spin-tensor
obeying $({\Sigma}{}^{i,  [3] ,[3']}\gamma^i)_{qp}=- {1\over 6}
\delta^{[k_1}_{j_1}\delta^{k_2}_{j_2}\delta^{k_3]}_{j_3}\delta_{qp}$.
As a consequence of the Rarita--Schwinger and Einstein equations of
the 11D supergravity, the above fluxes obey
\begin{eqnarray}\label{hatRS=} & \gamma^{i}_{qp}\hat{T}_{\# i \, +p}=0
\; \; (a) \; ,    \quad  \hat{R}_{\# j\# j} + {1\over 3} (\hat{F}_{\#
ijk})^2= 0  \;  \; (b) \; .\quad
\end{eqnarray}

{\bf 2.} We describe the {\it relative motion of} mM$0$ {\it constituents} by
the maximally supersymmetric {\it $SU(N)$ YM gauge theory on ${\cal
W}^{(1|16)}$} whose embedding into the target 11D superspace is
specified by the superembedding equation (\ref{SembEq=M0})
\cite{mM0=PLB}. This latter results in dynamical equations which
formally coincide with the single M$0$ equations and thus describes, in
terms of coordinate functions $\hat{Z}^{{ {M}}}(\zeta)= (\hat{x}{}^{
{m}}(\zeta)\, , \hat{\theta}^{\check{ {\alpha}}}(\zeta))$, the
motion of the center of energy of multiple M$0$--brane system. The
gauge theory is formulated in terms of 1-form gauge potential
$A=e^{\# }A_{\# }+ e^{+q} A_{+q}$  on ${\cal W}^{(1|16)}$. Its field
strength  $G_2= dA - A\wedge A= e^{\cal A}e^{\cal B}G_{{\cal B}{\cal
A}}/2$ obeys the constraints
\begin{eqnarray}\label{M0:G=sX}
G_{+q\, +p}=  i \gamma^i_{qp} {\bf X}{}^i \; , \qquad
\end{eqnarray}
where $\gamma^i_{qp}= \gamma^i_{pq}$ are nine-dimensional Dirac
matrices, $\gamma^i\gamma^j+ \gamma^j\gamma^i =\delta^{ij}I_{16\times 16}$ ($\; i=1,\ldots , 9$).
This constraint involves a nanoplet of N$\times$N hermitian matrix
superfields ${\bf X}{}^i =({\bf X}{}^i)^\dagger$ the leading
component of which provides the natural candidate for the field describing
the relative motion of the M0 constituents.

Studying Bianchi identities one finds that the  selfconsistency
of the constraints (\ref{M0:G=sX}) requires the
matrix superfield ${\bf X}{}^i $ to obey the {\it superembedding--like
equation} \cite{mM0=PLB}
\begin{eqnarray}\label{M0:DX=gP}
D_{+q}{\bf X}{}^i= 4i\gamma^i_{qp}{\mathbf \Psi}_{q}\; . \qquad
\end{eqnarray}

The set of physical fields of the $d$=1, ${\cal N}$=16 SYM model
defined by constraints (\ref{M0:G=sX}) is exhausted by the leading
component of the bosonic superfield ${\bf X}{}^i $, providing the
non-Abelian, $N\times N$ matrix generalization of the Goldstone
field describing a single M$0$-brane in static gauge, and by its
superpartner, the leading component of the fermionic superfield
${\mathbf \Psi}_{q}$ in (\ref{M0:DX=gP}), providing the non-Abelian,
$N\times N$ matrix generalization of the fermionic Goldstone field
describing a single M0-brane.
(which can be extracted from the
fermionic coordinate function of M0-brane by fixing the gauge with
respect to local fermionic $\kappa$--symmetry).
To be convinced in
that no other fields appear, one can calculate the spinor
covariant derivative of the fermionic superfield and find
\begin{eqnarray}\label{M0:Dpsiq=}
& D_{+p}{\mathbf \Psi}_{q}\; = {1\over 2}\gamma^i_{pq}D_{\#}{\bf
X}^i + {1\over 16} \gamma^{ij}_{pq} \; [{\bf X}^i, {\bf X}^j] -
\qquad \nonumber \\ & - {1\over 12} {\bf X}^i\hat{F}_{\#
jkl}\left(\delta^{i[j}\gamma^{kl]}+ {1\over 6}
\gamma^{ijkl}\right)_{pq}
 \; .
\end{eqnarray}

{\bf 3.} {\it Equations of motion and polarization of multiple M0 by
flux.} Studying the selfconsistency condition of Eq.
(\ref{M0:Dpsiq=}) (on the line of \cite{mM0=PLB} but
taking into account nonvanishing supergravity fluxes) we find the
interacting dynamical equation for the fermionic matrix (super)fields
\begin{eqnarray}\label{M0:DtPsi=}
& D_{\#}\Psi_{q}=- {1\over 4} \gamma^i_{qp} \left[ {\bf X}^i\, , \,
\Psi_{p} \right]   -  {1\over 24}  \hat{F}_{\# ijk}
\gamma^{ijk}_{qp}\Psi_{p}  - \nonumber  \\ & - {1\over 4}  {\bf
X}^i\hat{T}_{\# i \, +q}\, . \qquad
\end{eqnarray}
As usual in supersymmetric theories, the higher components in
decomposition of the superfield version of the fermionic equations
over the Grassmann coordinates of ${\cal W}^{(1|16)}$ give the bosonic equations of motion. In the case of our multiple M$0$ system these are the Gauss constraint
\begin{eqnarray}\label{M0:XiDXi=}
& \left[ {\bf X}^i\, , \, D_{\#}{\bf X}^i\,  \right] = 4i
\left\{{\mathbf \Psi}_{q}\, , \, {\mathbf \Psi}_{q} \right\} \;
\qquad
\end{eqnarray}
and proper equation of motion
\begin{eqnarray}\label{M0:DDXi=}
& D_{\#}D_{\#} {\bf X}^i = {1\over 16} \left[ {\bf X}^j\, , \,
\left[ {\bf X}^j\, , \, {\bf X}^i\,  \right]\right]+ i\gamma^i_{qp}
\left\{{\mathbf \Psi}_{q}\, , \, {\mathbf \Psi}_{p} \right\} +
\nonumber \\ &  + {1\over 8} \hat{F}_{\# ijk} \left[ {\bf X}^j\, ,
\, {\bf X}^k\, \right] +  {1\over 4}  {\bf X}^j \hat{R}_{\# j\; \#
i} -2i \Psi_{q}\hat{T}_{\# i \, +q} \; . \qquad
\end{eqnarray}
The third term in the {\it r.h.s.} of the bosonic equation
(\ref{M0:DDXi=}), $\hat{F}_{\# ijk} \left[ {\bf X}^j\, , \, {\bf
X}^k\,  \right]$, is essentially non-Abelian and typical for
`dielectric coupling' characteristic for the Emparan-Myers
`dielectric brane effect' \cite{Emparan:1997rt,Myers:1999ps}. The
fourth term is the mass term for  $N\times N$
matrix $SO(9)$ vector superfield ${\bf X}^j$ with the
mass matrix given by the projection of Riemann tensor
$\hat{R}_{\#i\; j\#}$ (=$\hat{R}_{\#j\; i\#}$) defined in  Eq. (\ref{M0:RFlux}).

{\bf 4.} Actually, using only the SO(1,1)$\times$SO(9) symmetry of our mM$0$ system one can not only find all the terms in the {\it r.h.s.}'s of Eqs. (\ref{M0:DtPsi=})--(\ref{M0:DDXi=}), but also conclude that only two other contributions might be possible but are absent. The reason beyond this {\it rigid structure of the multiple M$0$ equations} is that all the basic superfields and projections of the background fluxes interacting with mM$0$ constituents carry positive $SO(1,1)$ weights ('charges').

Indeed, the $SO(1,1)$ weights are $+2$ for the bosonic superfield  ${\bf X}^i
:={\bf X}^i_{++}$, $+3$ for the fermionic ${\mathbf
\Psi}_{q}:={\mathbf \Psi}_{++ \, +q}$, and $+2$, $+3$ and
$+4$, respectively, for the 11D supergravity fluxes  $\hat{F}_{\# jkl}:=\hat{F}_{++ jkl}$
(\ref{M0:Fluxes}), $\hat{T}_{\# j \, +p}:=\hat{T}_{++ j \, +p}$
(\ref{M0:fFluxe}) and $\hat{R}_{\#i\; j\#}:= \hat{R}_{++\, i\; j\,
++}$ (\ref{M0:RFlux}). Then, as the covariant derivative $D_\#:=
D_{++}$ has the weight $+2$, the fermionic and bosonic
equation of motion are N$\times$N matrices
with the SO(1,1) weights $+5$ and $+6$, respectively. Now, taking into
account also the  $SO(9)$ index structure of the basic superfields and fluxes,
one sees that, if we do not allow ourselves using the inverse and fractional powers
of  $\hat{R}_{\#i\; i\#}$ and of matrix
superfield ${\bf X}^i{\bf X}^i$, very few terms can be written in
addition to ones already present in (\ref{M0:DtPsi=})-(\ref{M0:XiDXi=}). Moreover, all but two of these actually vanish.

Indeed, one could add
${\bf X}{}^i\gamma^{ij}T_{\# j + p}$ to the Dirac equation (\ref{M0:DtPsi=}), however  this term can be expressed through the already present ${\bf X}{}^iT_{\#
i + p}$ using (\ref{hatRS=}a). This equation is also
responsible for vanishing, ${\mathbf
\Psi}_q\gamma^{j}_{qp}T_{\# j +p}=0$, of the only possible contribution to Eq.  (\ref{M0:XiDXi=}), and for that the
possible fermionic contribution ${\mathbf \Psi}_q\gamma^{ij}T_{\# j
+ p}$ to Eq. (\ref{M0:DDXi=}) can be expressed in terms of ${\mathbf
\Psi}_qT_{\# i + q}$ already present there. As far as the pure
bosonic contributions to (\ref{M0:DDXi=}) are concerned, the already
present terms could be completed by the  $
{\bf X}^i\hat{R}_{\# k\; \# k}$ and ${\bf X}^j \hat{F}_{\#
ikl}\hat{F}_{\# jkl}$ (due to (\ref{hatRS=}b), $\hat{F}_{\# jkl}\hat{F}_{\# jkl}\,
\propto \,\hat{R}_{\# k\; \# k}$). Thus the only results of
the explicit calculations in the frame of superembedding approach are the absence of these two contributions
to the mass matrix of the N$\times$N matrix superfield ${\bf X}^i$
and the exact values of the nonvanishing coefficients in (\ref{M0:DtPsi=})-(\ref{M0:DDXi=}).

Such a rigid structure of the mM$0$ equations (\ref{M0:DtPsi=})--(\ref{M0:DDXi=}), which suggests their universality, comes from SO(1,1)$\times\,$SO(9) symmetry of our mM$0$ system. This originates in that M$0$-brane is actually the {\it massless} 11D superparticle the momentum of which is light--like and has a small group which is essentially SO(1,1)$\times$SO(9)  (see \cite{IB07:M0} and refs. therein for a more precise statement). Such a rigidity cannot be seen from observing the mD$0$ equations
\cite{IB09:D0} as they are: as D$0$-brane is a massive 10D superparticle, the (spacial) symmetry of the relative motion of mD$0$ system of \cite{IB09:D0} is restricted to SO(9), the small group of the timelike 10D momentum. Thus, to see the rigid structure of mD$0$ equations \cite{IB09:D0}, one needs to appreciate their 11D origin, {\it i.e.} their appearance as a result of dimensional reduction of our mM$0$ equations.

{\bf 5.} The {\it BPS conditions} for the supersymmetric bosonic
solutions of Eqs. (\ref{M0:DDXi=}) and (\ref{M0:XiDXi=}) can be
obtained from Eqs. (\ref{DpT--iq=}) and (\ref{M0:Dpsiq=}). For 1/2
supersymmetric configurations these (1/2 BPS equations) read
\begin{eqnarray}\label{1/2BPSeq=0} & D_{+p} \hat{T}_{\# \, i\, +
q}\vert_{0}= 0 \; \;   (a) \; , \quad D_{+p}\Psi_q\vert_{0}=0 \; \; (b)
\; . \quad
\end{eqnarray}
Eq. (\ref{1/2BPSeq=0}a) restricts the 3-form flux pull--back to be
constant, $D_\# \hat{F}_{\# ijk}=0$,  and, up to $SL(9)$
transformations, to have the form $\hat{F}_{\# ijk}= 3/4f \delta^i_I
\delta^j_J \delta_K^k \epsilon^{IJK}\;$ ($I=1,2,3$). Then Eq.
(\ref{1/2BPSeq=0}b) has the fuzzy 2--sphere solution
\begin{eqnarray}\label{M0:BPS=fuS3}
 {\bf X}^i = \delta^i_I  f T^I\; , \qquad
 {}[  T^I , T^J ] = \epsilon^{IJK} T^K
 \; . \qquad
\end{eqnarray}
This configuration was known to solve the pure bosonic equations of \cite{Myers:1999ps}, but in our case it appears as describing the M$2$--brane as a
supersymmetric configuration of mM$0$ system, and also the relation of the special form of the flux ($\hat{F}_{\# ijk}\propto  \epsilon^{IJK}$) with preservation of 16 supersymmetries becomes manifest. Curiously, the famous Nahm equation $D_\# {{\bf X}}{}^I + {i\over 8}
\epsilon^{IJK}  [{\bf X}^J, {\bf X}^K] =0$ \cite{NahmEq}, which also has fuzzy--two--sphere--related solution,   appears as an SO(3) invariant 1/4 BPS condition for the case of {\it vanishing} 4-form flux
pull--back, $\hat{F}_{\# ijk}=0$.

{\bf 6.} Giving a covariant and supersymmetric description of the Matrix model interaction with nontrivial 11D supergravity fluxes, our approach might provide a new framework for studying M--theory. The first of the promising directions is to search for other supersymmetric solutions of the mM0 equations, representing more complicated M-brane and D-brane configurations.

For the development of our approach it is important  to clarify whether our superembedding description can be generalized for multiple M$p$-branes and D$p$-branes with higher $p$. An important problem is also to find an
action functional for the embedding functions and matrix
superfields ${\mathbf X}^i$, ${\mathbf \Psi}_q$ which reproduced our mM$0$ equations.
To this end the application of the 'Ectoplasm--like'
technique of restoring the action from superembedding approach (see
\cite{Howe+Linstrom+Linus} and refs. therein) looks promising. Another challenge is to understand whether one can develop a counterpart of the (string-inspired and hence seemingly ten dimensional) boundary fermion approach \cite{Howe+Linstrom+Linus} for the eleven dimensional multiple M0--system.

The author is thankful to Dima Sorokin, Mario Tonin and Paolo Pasti
for the interest to this work and for their warm hospitality at the
Padova Section of INFN and Padova University on its final stages.
The partial support  by the research grants FIS2008-1980 from the
Spanish MICINN  and 38/50--2008 from Ukrainian Academy of
Sciences and Russian RFFI is greatly acknowledged.


\begin{thebibliography}{99}
\renewcommand{\theequation}{R.\arabic{equation}}

\bibitem{M-theory}
C.M.~Hull and P.~K.~Townsend,
Nucl. Phys. {\bf B438}, 109 (1995);
E.~Witten,
Nucl. Phys.  {\bf B443}, 85 (1995);
J.H.~Schwarz,
{Nucl.Phys.Proc.Suppl.} {\bf B55}, 1 (1997);
 P.K.~Townsend, {\it
Four Lectures on M--theory}, hep-th/9612121. In: {\sl Summer School
High energy physics and cosmology, Trieste 1996} (Eds. E. Gava, et
al)
World Scientific, 1997, Singapore,
pp. 385-438 [hep-th/9612121].


\bibitem{AdS}
J.M. Maldacena, {Adv. Theor.Math.Phys.} {\bf 2}, 231(1998);
\, S.S. Gubser, I.R. Klebanov, A.M. Polyakov, {Phys. Lett.}
 {\bf B428}, 105 (1998);
\, E. Witten, {Adv. Theor.Math.Phys.} {\bf 2}, 253 (1998);
\, O. Aharony, S.S. Gubser, J. Maldacena, H. Ooguri, Y. Oz,
{Phys.Rept.} {\bf  323}, 183 (2000).

\bibitem{Randall:1999ee}
  L.~Randall and R.~Sundrum,
  Phys.\ Rev.\ Lett.\  {\bf 83}, 3370 (1999);
  Phys.\ Rev.\ Lett.\
  {\bf 83}, 4690 (1999).

\bibitem{Duff:2000az}
  M.~J.~Duff, J.~T.~Liu and K.~S.~Stelle,
  J.\ Math.\ Phys.\  {\bf 42}, 3027 (2001)
  [arXiv:hep-th/0007120].


\bibitem{Duff94-Stelle98}
  M.~J.~Duff, R.~R.~Khuri and J.~X.~Lu,
  Phys.\ Rept.\  {\bf 259}, 213 (1995)
  [hep-th/9412184];
  K.~S.~Stelle,
  ``BPS branes in supergravity,''
in:
{\it High Energy Physics and Cosmology  1997} (Eds: E. Gava et al.),
Singapore, World Scientific, 1998,
pp. 29-127 [hep-th/9803116].

\bibitem{BST1987}
  E.~Bergshoeff, E.~Sezgin and P.~K.~Townsend,
  Phys.\ Lett.\  {\bf B189}, 75 (1987).

\bibitem{Dpac}
P.~K.~Townsend,
  Phys.\ Lett.\  {\bf B373}, 68 (1996)
  [hep-th/9512062];
M.\ Cederwall, A.\ von Gussich, B.E.W.\ Nilsson, A.\ Westerberg,
{Nucl.Phys.} {\bf B490}
(1997) 163  [hep-th/9610148]; 
M. Cederwall, A.\ von Gussich, B.E.W.\ Nilsson, P.\ Sundell and A.\
Westerberg,
{Nucl.Phys.} {\bf B490}, 179 (1997)
[hep-th/9611159]; M.\ Aganagic, C.\ Popescu, J.H.\ Schwarz,
{Phys.Lett.} {\bf B393} (1997) 311 [hep-th/9610249];
{Nucl.Phys.} {\bf B490}, 202 (1997) [hep-th/9612080].

\bibitem{B+T=Dpac}
E.\ Bergshoeff, P.K.\ Townsend,
{Nucl.Phys.}
{\bf B490}, 145--162 (1997)  [hep-th/9611173].

\bibitem{blnpst}
  I. Bandos, K. Lechner, A. Nurmagambetov, P.~Pasti, D.~Sorokin, M.~Tonin,
  Phys. Rev. Lett. {\bf 78},  4332(1997)
  [hep-th/9701149];
M.~Aganagic, J.~Park, C.~Popescu, J.H. Schwarz,
  Nucl. Phys. {\bf B496}, 191 (1997)
  [hep-th/9701166].


\bibitem{bpstv}
  I. Bandos, D. Sorokin, M. Tonin, P. Pasti, D.V. Volkov,
  Nucl. Phys.  {\bf B446}, 79-118 (1995)
  [hep-th/9501113].



\bibitem{hs96}
 P.S.~Howe and E.~Sezgin,
  Phys. Lett.  {\bf B390}, 133 (1997)
  [hep-th/9607227].
\bibitem{hs2}
  P.S.~Howe and E.~Sezgin,
  Phys.\ Lett.
{\bf B394}, 62 (1997)
  [hep-th/9611008].




\bibitem{Dima99}
  D.~P.~Sorokin,
  Phys.\ Rept.\  {\bf 329}, 1-101 (2000).

\bibitem{IB09:D0}
  I.A. Bandos,
  Phys.\ Lett.\  {\bf B680}, 267-273 (2009).

\bibitem{mM0=PLB}
  I.A. Bandos,
  Phys.\ Lett.\  {\bf B687}, 258--263 (2010).

\bibitem{Witten:1995im}
  E.~Witten,
  Nucl.\ Phys.\   {\bf B460}, 335 (1996).

\bibitem{BLG}
  J.~Bagger and N.~Lambert,
  Phys.\ Rev.\  {\bf D77}, 065008 (2008);
  JHEP {\bf 0802} (2008) 105;
  A.~Gustavsson,
  Nucl.\ Phys.\  B {\bf 811}, 66 (2009).

\bibitem{ABJM}
  O.~Aharony, O.~Bergman, D.~L.~Jafferis and J.~Maldacena,
  JHEP {\bf 0810}, 091 (2008)
  [arXiv:0806.1218 [hep-th]].


\bibitem{Myers:1999ps}
  R.~C.~Myers,
  JHEP {\bf 9912}, 022 (1999)
  [hep-th/9910053].



\bibitem{YLozano+=0207}
 B.~Janssen and Y.~Lozano,
  Nucl.\ Phys.\  {\bf B643}, 399 (2002);
  [hep-th/0205254];
  {\bf B658}, 281 (2003)
  [hep-th/0207199].

\bibitem{Emparan:1997rt}
  R.~Emparan,
  Phys. Lett.  {\bf B423}, 71 (1998).




\bibitem{Howe+Linstrom+Linus}
 P.~S.~Howe, U.~Lindstrom and L.~Wulff,
  JHEP {\bf 0508}, 041 (2005);
  JHEP {\bf 0702}, 070 (2007)
  [hep-th/0607156].




\bibitem{Dima01}
 D.~Sorokin,
  JHEP {\bf 08}, 022 (2001);
  J.~Drummond, P.S.~Howe and U.~Lindstrom,
  Class.Quant.Grav.  {\bf 19}, 6477 (2002);
S.~Panda and D.~Sorokin,
  JHEP {\bf 02},  055 (2003).


\bibitem{Banks:1996vh}
  T.~Banks, W.~Fischler, S.~H.~Shenker and L.~Susskind,
  Phys.\ Rev.\  {\bf D55}, 5112-5128 (1997)
  [hep-th/9610043].

\bibitem{CremmerFerrara80BrinkHowe80}
E.~Cremmer and S.~Ferrara,
Phys. Lett. {\bf B91}, 61 (1980);
L.~Brink and P.S.~Howe,
Phys. Lett. {\bf B91}, 384 (1980).


\bibitem{stv}
  D.~P.~Sorokin, V.~I.~Tkach and D.~V.~Volkov,
  Mod.\ Phys.\ Lett.\  {\bf A4} (1989) 901-908.

\bibitem{IB07:M0}
  I.A. Bandos,
  Nucl. Phys. {\bf B796}, 360 (2008).

\bibitem{NahmEq}
  W.~Nahm,
  Phys.\ Lett.\  {\bf B90}, 413 (1980).

\end{thebibliography}
\end{document}